\documentclass[aps,preprint,floatfix]{revtex4}
\usepackage{amsmath,units,xspace,amssymb}
\usepackage[pdftex]{graphicx}
\usepackage{booktabs}
\begin{document}

%\linenumbers

\title{Probing the Local Response of Glass-forming Liquids by Laser Excitations}
\author{Bo Li,$^{1,*}$ Kai Lou,$^{1,*}$ Walter Kob$^{2, \ddagger}$ and Steve Granick$^{1,\dagger}$}
\affiliation{1, Center for Soft and Living Matter, Institute of Basic Science, Ulsan, 44919, South Korea \\2, Laboratoire Charles Coulomb (L2C), University of Montpellier, CNRS, Montpellier, France}
\date{\today}

\begin{abstract} % <150 words

Glass is a disordered solid that processes distinct dynamical and elastic properties compared with crystal. How heterogeneous glassy materials can be and to what extend dynamics is encoded with structure and elasticity are long standing puzzles in glass science. In this experiment, we probed the responses of binary colloidal glasses towards the excitations induced by highly focused laser pulses. We observed very similar excitation patterns when the laser was repeated in linear region; directly proving that the dynamical heterogeneity is strongly encoded with structure. In non-linear region, we identified a non-monotonic dynamical length scale as a function of area fraction, resulting from non-monotonic coupling of momentum transfer in radial and orthogonal directions. Surprisingly, the excitation size and radius of gyration conformed to a universal scaling relation that covered both linear and non-linear regions. Our experiments offered a new strategy of actively probing the response of glassy materials on the microscopic level.

\end{abstract}

\maketitle

\textbf{$*$: these authors contributed equally to this work.}

\textbf{corresponding email: $\ddagger$ walter.kob@umontpellier.fr, $\dagger$ sgranick@ibs.re.kr}

\newpage

In contrast with crystal, glassy materials are characterized by a lack of long-range order, whether at the atomic level or at much larger length scales \cite{2011RMP_Berthier}. Characteristic features of glassy dynamics, including dynamical heterogeneity (DH) \cite{1997PRL_Kob}, that appear below an onset temperature are therefore qualitatively captured by mean field theory \cite{2015RMP_Kirkpatrick}. However, the spatially extending disorder does not necessarily ensure the microscopically uniform structure. The diversity of local structure (LS) \cite{2008NM_Royall} will naturally consequent in heterogeneity of local elasticity which is not fully considered by theories \cite{2006RMP_Dyre}. Structurally or elastically, how heterogeneous glassy materials can be \cite{2015N_Ketov, 2016S_Albert, 2016NM_Gelin} and to what extend structure and elasticity can influence DH \cite{2008NM_Royall, 2008NP_Cooper, 2012NP_Kob, 2016NP_Golde, 2017S_Cubuk} are long standing puzzles in glass science that lack decisive answer in experiment.

Whether DH is encoded with structure has long been under debate \cite{2008NM_Royall, 2008NP_Cooper, 2012NP_Kob, 2016NP_Golde, 2017S_Cubuk}. Due to its mean-field nature, the random first order transition (RFOT) theory is not at the position to offer a clear prediction on the relationship between LS and DH \cite{2015RMP_Kirkpatrick}. Molecular dynamics simulations showed weak spatial coupling between the soft spot identified by vibrational mode \cite{2008NP_Cooper, 2011PRL_Manning}, or LS \cite{2018PRX_Tong} with the highly mobile region (HMR); while kinetic theories believe DH is purely thermal fluctuation without any connection to structure \cite{2017PRX_Royall, 2017PRL_Cates}. There are also some indirect links between structural and dynamics in scale of ensemble averaged quantities \cite{2017JSMTE_Royall, 2018PRL_Wang}. The lack of conclusive prove probably comes from the spontaneous nature of HMR. In passive studies without external perturbation, the HMRs various quite randomly in space and time. The simultaneous change in both dynamics and structure hinders the `in situ' observation of dynamical relaxation at fixed positions.

As the temperature of a glass-forming liquid is decreased, three characteristic points appear successively. They are, from higher to lower temperature, the point at which the caging takes into effect ($T_\textrm{onset}$) \cite{1995S_Angell}, the mean field transition point ($T_\textrm{MCT}$) \cite{1997arXiv_Kob} and the ideal glass transition point ($T_\textrm{g}$) \cite{1995S_Angell}. Although, guided by theories \cite{2011RMP_Berthier}, many studies have focused on the DH in the deeply superooled regime \cite{2008NM_Royall, 2008NP_Cooper, 2012NP_Kob, 2016NP_Golde, 2017S_Cubuk}, i.e. close to $T_\textrm{MCT}$ or $T_\textrm{g}$, the properties near $T_\textrm{onset}$ were broadly overlooked in glass science. However, at $T_\textrm{onest}$, significant changes were observed in both structural and dynamical aspects. For example, certain motifs start to occur frequently and the mean square displacement exhibits a shoulder at this point. It's therefore reasonable to expect that there is also a signature in the mechanical susceptibility. To reveal the mechanical properties, it also requires `in situ' observation of local excitation and response simultaneously.

Colloids have served as outstanding model systems for the study of glass transitions \cite{2013ARCMP_Lu}. Micro particles dispersed in water undergo Brownian motion, which perfectly simulates the diffusion of atoms or molecules \cite{2016NRM_Li}. The motion of each particle can be recorded and digitalized using optical video microscopy \cite{1996ARPC_Murray}. Hence, kinetic information with single$\textrm{-}$particle resolution can always be obtained on these platforms. In combination with techniques like optical tweezers \cite{2015NP_Nagamanasa} and shear \cite{2009N_Mattsson}, the interplay between glassy materials and external perturbation has been investigated. Recently, the response of 2D colloidal crystal has been tested using optical laser beams \cite{2017PNAS_Buttinoni, 2018PRL_Cash, 2018PNAS_Lavergne}. In particular, the excitation experiments through local laser pulses exhibited perfect compliance to the classical elastic theory \cite{2017PNAS_Buttinoni}. These laser excitation methods \cite{2017PNAS_Buttinoni, 2018PRL_Cash, 2018PNAS_Lavergne} offer the opportunity of locally and actively probing the dynamical and elastic properties of colloidal glasses.

\vspace{12pt}
\textbf{The local excitation method}

In this manuscript, we report the first experimental test of the local responses of colloidal glass forming liquids using laser excitation. By shining local and short-period laser pulse ($0.5~s$) to the colloidal samples, we were able to transfer the energy of light to the colloidal particles (Figs.~1, S1, S2) in the form of kinetic energy and thus actively generate HMR, i.e., local excitation event. When the laser was on at $t=0$, the particles around the laser spot were immediately kicked away from their original position (Movie~1). The displacement of the most nearby particles reached a platform $0.3~$s after the laser was on (Fig.~1b). The mobility of outer layered particles was then subsequently enhanced by the collisions from inner ones. After the laser was off at $t=0.5~$s, the particles either stayed at the new positions (Fig.~1b) or relaxed back to their original positions (Fig.~S2a), depending on the local elasticity/plasticity and the strength of the external perturbation. The two-dimensional colloidal samples consisted of binary spherical particles that were confined between two glass walls. The number and size ratio between the small and large particles were chosen basing on the Kob-Anderson model \cite{1995PRE_Kob}. The onset packing fraction ($\phi_\textrm{onset}$) was determined to be 0.60 by mean square displacement data (Fig.~S3a) and non-Gaussian parameter (Fig.~S3b); while the mean field transition point ($\phi_\textrm{MCT}$) was greater than 0.75 according to the intermediate scattering function (Fig.~S3c).

\begin{figure}[!h]
	\centering
	\includegraphics[width=1.0\columnwidth]{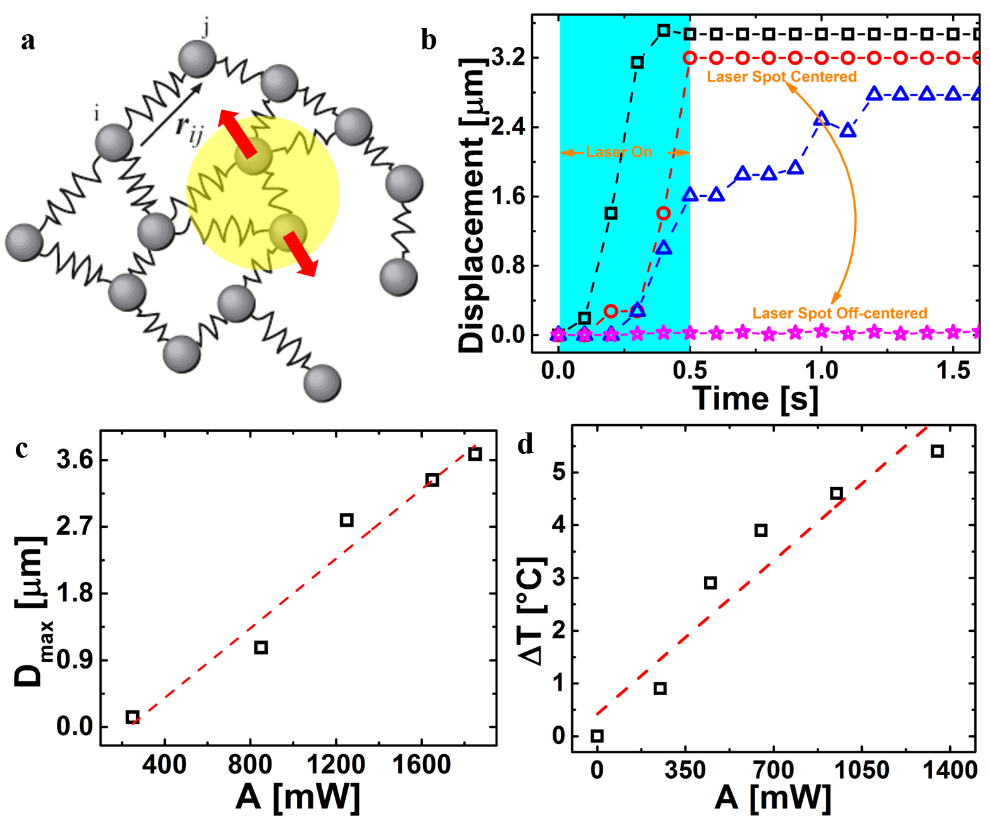}
	\caption{\textbf{The experimental system.} \textbf{a}, The schematic of a laser excitation event within colloidal sample. \textbf{b}, Displacements of particles near an fixed high power laser spot after it was on at $t=0~s$ in a sample with $\phi=0.45$. \textbf{inset}, The linear relationship between the maximum displacement and $A$.}
	\label{fig:1}
\end{figure}

For each laser strength ($A$) and the area fraction ($\phi$), we started to monitor the relaxation of the system right after the laser was on ($t=0~s$). By selecting out the particles whose displacement exceeded a threshold value of 0.3~$\sigma_\textrm{s}$ during $\Delta t=5~s$, we quantified the size of local excitation ($S=\dfrac{\pi}{4}(n_\textrm{s} \sigma_\textrm{s}^{2}+n_\textrm{l} \sigma_\textrm{l}^{2})$) as a function of $t$. Here, $\sigma_\textrm{s}$ and $\sigma_\textrm{l}$ was the diameters of the small and large particles respectively; while $n_\textrm{s}$ and $n_\textrm{l}$ was the number of small and large particles whose displacement satisfied the selecting criteria. We found $S$ reached a maximum value $S_\textrm{max}$ at about $t=0.7-2.0~s$ for different $\phi$ values (Fig.~S2b-e), approximately $0.2-1.5~s$ after the laser was off. We then used this $S_\textrm{max}$ as the excitation size under the given ($\phi,~A$). Note that $\Delta t=5~s$ was much smaller than the relaxation time measured from the intermediate scattering function (Fig.~S3c), ensuring no crossing time scales. The similar shape of radial distribution function for different $\phi$ (Fig.~S3d) suggested no structural instability in our experiments. More details about the experimental system and quantification of $S_\textrm{max}$ can be found in the \textbf{Methods} section.

Basing on the $S_\textrm{max}(\phi, A)$ data, we investigated the responses of the colloidal glasses in both linear (small $A$) and non-linear (large $A$) regions. The analysis of the dynamical length scale (DLS) and the morphology of excitations established direct links between the dynamics and structure that are probably general for glassy materials.

\vspace{12pt}
\textbf{Dynamical heterogeneity encoded with local structure and elasticity} 

When a beam of laser pulse was shined to a doped gold particle, it absorbed the heat effectively and imposed strong scattering forces to its neighbors \cite{2017PNAS_Buttinoni}. A HMR was thus created through the collisions among particles. We visualized the relaxation of HMR, i.e., the excitation pattern, by coloring the particles' displacements within 5~$s$ after the laser was off (Fig.~2a-c). We then repeatedly shined the laser to the gold particle with time separations much longer than the relaxation time of the sample, making sure that the sample was fully in equilibrium before each excitation event.

\begin{figure}[!h]
	\centering
	\includegraphics[width=1.0\columnwidth]{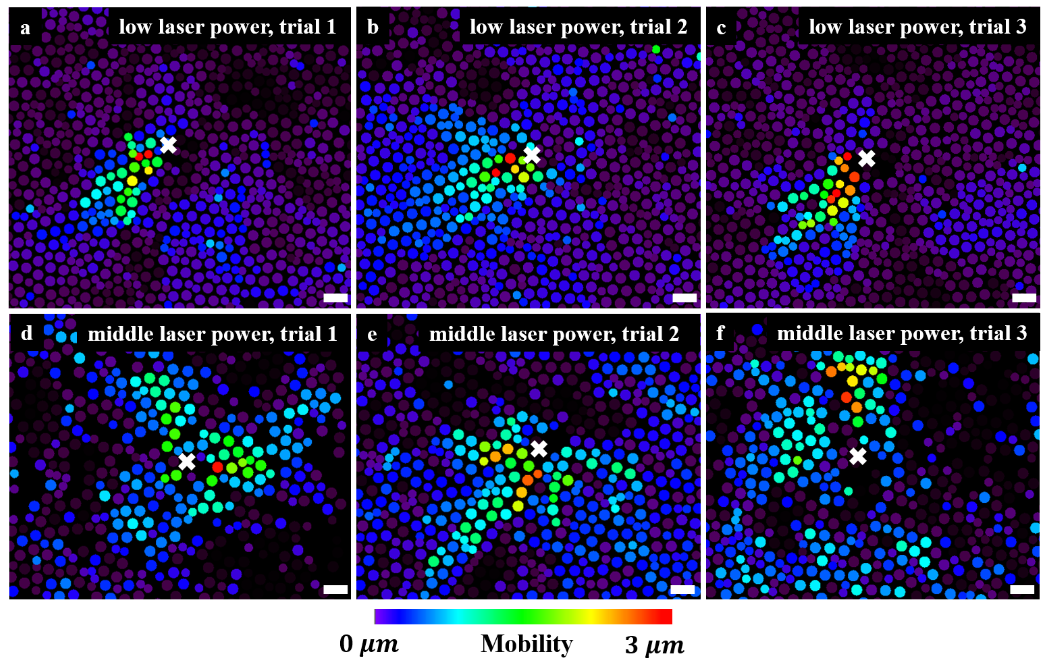}
	\caption{\textbf{Dynamics encoded with structure and elasticity.} \textbf{a-c}, Excitation patterns in linear region. An low power laser pulse was repeatedly shined to a doped gold particle (marked by the white cross) in a $\phi=0.58$ sample. \textbf{d-f}, Excitation patterns in non-linear region. An middle power laser pulse was repeatedly shined to a doped gold particle (marked by the white cross) in a $\phi=0.58$ sample. The color represents the particles' displacements within 5~$s$ at $t=2.0~s$. The duration of the pulse was 0.5~$s$. The separations between two excitation events were 20~$min$. The scale bar is 10~$\mu m$.}
	\label{fig:2}
\end{figure}  

It's unambiguously shown in Fig.~2a-c that the relaxations of HMR were very similar to each other for different trials when the laser was well in linear region (lower laser power) (Fig.~2a-c). In this case, no irreversible rearrangement among particles occurred and the LS was exactly the same before each excitation. However, when we increased $A$ to the level that were strong enough (middle laser power) to alter the LS, completely different excitation patterns were observed (Fig.~2d-f). The outcome from the `gold-particle' method was further confirmed by that of the `four-point' method (Figs.~S5, S6), in which a single beam of light was further splitted into four so that we could simultaneously monitor the responses of different positions within the same sample. These observations clearly proved that the DH, or more specifically, the relaxation of HMR were strongly encoded with LS in linear region. 

In addition, a positive relation between $S_\textrm{max}$ and local elastic modulus ($\sigma_\textrm{xy}$) existed in both linear and non-linear regions (Fig.~S7a). Here, $\sigma_\textrm{xy}$ was measured by `stress assessment from local structural anisotropy (SALSA)' method \cite{2016NM_Lin}, which was proved to be highly accurate for quantifying the elasticity of liquids \cite{2017PRL_Lin} and crystals \cite{2016NM_Lin} composed of hard-sphere-like colloidal particles. Regardless of $\phi$ (size of data points in Fig.~S7a) and $A$, the $S_\textrm{max}$ was solely determined by the local elasticity of the position at which the external perturbation was imposed. We know that $S_\textrm{max}$ reflected the extent to which the relaxation of a HMR can reach and thus was an indicator of DH; while according to SALSA method, $\sigma_\textrm{xy}$ was eventually determined by the LS. Therefore, the positive correlation between $S_\textrm{max}$ and $\sigma_\textrm{xy}$ was a direct link between dynamics and structure. The fact that the data points in Fig.~S7a included excitation events in both linear and non-linear regions generalized the conclusion that DH is encoded with LS \cite{2008NP_Cooper, 2004PRL_Cooper} to the whole $\phi$~-~$A$ phase diagram of colloidal glass. 

\vspace{12pt}
\textbf{Non-monotonic dynamical length scale in non-linear region} 

We observed a non-monotonic DLS in non-linear region. As shown in Fig.~3a, b, when $A$ reached a threshold value (800~$mW$), both averaged $S_\textrm{max}$ and the variance of $S_\textrm{max}$ ($\sigma (S_\textrm{max})$) exhibited a peak ($\phi_\textrm{max}$) close to $\phi_\textrm{onset}$. This non-monotonic behavior indicated that both the DLS and degree of DH reached the maximum at $\phi_\textrm{max}$. The non-monotonic behaviors disappeared at weak driving (Figs.~3, S10), when the system was well in elastic region. `Four-point' method was adopted for quantifying the excitations since it was able to offer four times the statistics as the `gold-particle' method.

Analysis on the particles' displacement revealed that the excited particles moved most cooperatively at $\phi_\textrm{max}$. For those particles that belong to the excitations, we decomposed their displacement within 5~$s$ to radial and orthogonal parts, referring to the position of the laser spot. We then calculated the ratio between the radial displacement and the total displacement, and plotted this participation ratio ($d_\textrm{r}/d$) as a function of $\phi$ in Fig.~4a. In accordance with $S_\textrm{max}$ (Fig.~3a) and $\sigma (S_\textrm{max})$ (Fig.~3b), $d_\textrm{r}/d$ peaked near $\phi_\textrm{onset}$ (Fig.~4a). Note that larger $d_\textrm{r}/d$ indicated higher momentum transfer efficiency in the radial direction, which was the direction of the driving force. We can therefore conclude that at $\phi_\textrm{max}$, the excited particles moved most cooperatively in response of the driving force, leading to largest $S_\textrm{max}$ and $\sigma (S_\textrm{max})$.

\begin{figure}[!h]
	\centering
	\includegraphics[width=1.1\columnwidth]{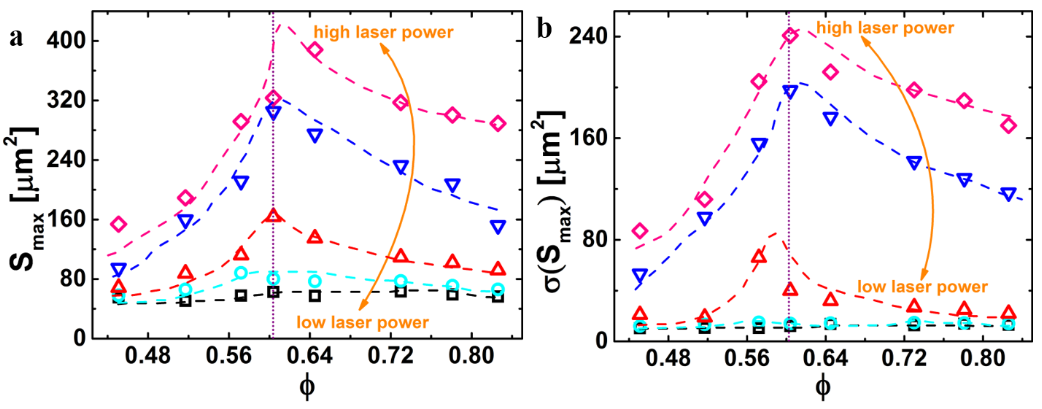}
	\caption{\textbf{Non-monotonic dynamical length scale.} \textbf{a}, The averaged excitation size as a function of $\phi$. \textbf{b}, The averaged variance of excitation size as a function of $\phi$. The vertical dotted line represented the onset area fraction, $\phi_\textrm{onset}$.}
	\label{fig:3}
\end{figure}

The fact that $\phi_\textrm{max}$ was near $\phi_\textrm{onset}$ instead of $\phi_\textrm{MCT}$ highlight the role $\phi_\textrm{onset}$ played during glass transition. When $\phi<\phi_\textrm{onset}$, the response of the system was in the `trivial' radial form. When $\phi>\phi_\textrm{onset}$, the response became more complicated since more fractal like excitations occurred. This observation implied that the system had a different local elastic behaviors that involved structural heterogeneity in high $\phi$ region in comparison with the simple continuous materials. Therefore, both the elastic and dynamical heterogeneities were deeply rooted in the onset of cage effect in glassy materials and can be reflected by local response. In this sense, the local response can be used to determine $\phi_\textrm{onset}$ without accessing the whole $\alpha$-relaxation process.

\vspace{12pt}
\textbf{Morphology of the Excitations}

To quantify the morphology of the excitations, we calculated the radius of gyration of them ($R_\textrm{g}$) using the formula $R_\textrm{g}=(\sum_{i=1}^{n} m_\textrm{i} r_\textrm{i}^{2}/\sum_{i=1}^{n} m_\textrm{i})^{1/2}$, where $m$ was the mass of the particle, $n$ was the number of particles belonging to the excitation and $r_\textrm{i}$ was the distance between the positions of the particle and laser spot. In accordance with $S_\textrm{max}$ (Fig.~3a), $\sigma (S_\textrm{max})$ (Fig.~3b) and $d_\textrm{r}/d$ (Fig.~4a), $R_\textrm{g}$ exhibited non-monotonic $\phi$-dependence that peaked near $\phi_\textrm{onset}$ (Fig.~4b). Note that $d_\textrm{r}/d$ was an indicator of momentum transfer efficiency in radial direction. The common peak of $d_\textrm{r}/d$ (Fig.~4a) and $R_\textrm{g}$ (Fig.~4b) revealed the fact that the relaxation and morphology of the excitations were mutually determined. Moreover, Fig.~4b showed that $R_\textrm{g}$ scattered only weakly at small $\phi$. After reaching a maximum near $\phi_\textrm{onset}$, the variance of $R_\textrm{g}$ (Fig.~4b), as well as the variance of $S_\textrm{max}$ (Fig.~3b), decayed again in high $\phi$ region. These observations (Figs.~3b, 4b) suggested that the system was most susceptible to the external perturbations near $\phi_\textrm{onset}$, implying a `transition' in the sense of mechanical responses.

\begin{figure}[!h]
	\centering
	\includegraphics[width=1.0\columnwidth]{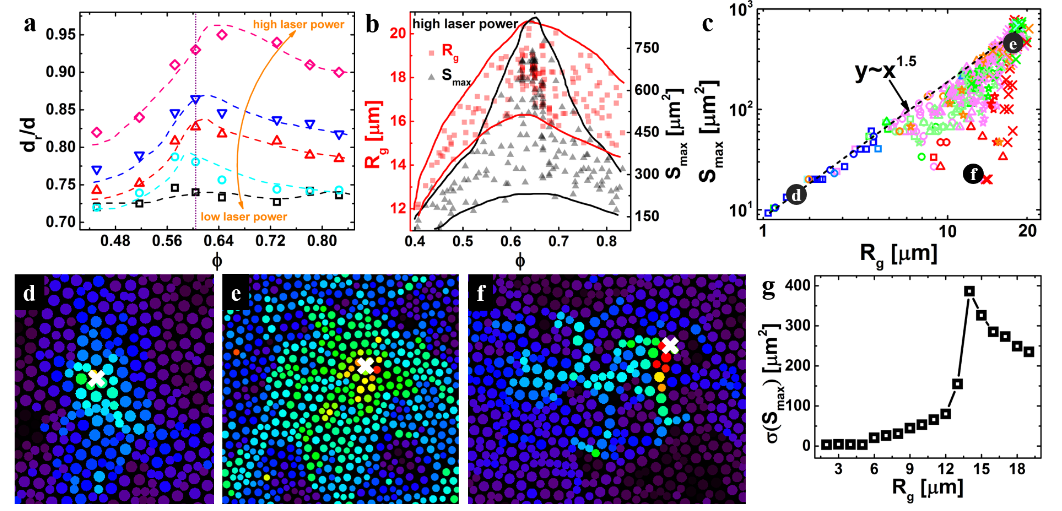}
	\caption{\textbf{Morphology of the excitations.} \textbf{a}, The averaged participation ratio of radial displacement as a function of $\phi$. The vertical dotted line represented the onset area fraction, $\phi_\textrm{onset}$. \textbf{b}, The radius of gyration and excitation size as a function of $\phi$ that include all data points at a fixed high laser power. \textbf{c}, The excitation size as a function of radius of gyration ($R_\textrm{g}$). The color and size of the data points indicated $A$ and $\phi$ respectively. The dotted line is the power-law fitting for the data points with $R_\textrm{g}<10$. The round markers \textbf{d-f} corresponded with the typical excitation pattern in \textbf{d-f}. \textbf{g}, the averaged variance of excitation size as a function of $R_\textrm{g}$.}
	\label{fig:4}
\end{figure}

Interestingly, $S_\textrm{max}$ and $R_\textrm{g}$ conformed to a scaling relation (Fig.~4c) that covered both linear and non-linear regions. The upper `envelop' of the data points can be adequately fitted into `power-law' growth, $y\sim x^{\alpha}$ with $\alpha=1.52 \pm 0.06$ (Fig.~4c). The fact that Fig.~4c included all data points regardless of $\phi$ and $A$ implied that the scaling was universal. The exponent $\alpha=1.52 \pm 0.06$ was a character of 1D procedures. When the transfer of momentum was radial, the relaxation of excitations was essentially a 1D process. Therefore, the universal scaling relation in Fig.~4c was a direct consequence of the nature of the excitations.

Another feature in the $S_\textrm{max}-R_\textrm{g}$ plot (Fig.~4c) was that the morphology heterogeneity occurred at middle length scale. We observed that $\sigma (S_\textrm{max})$ peaked at $R_\textrm{g}=14~\mu m$ (Fig.~4g) contributed by the scattered data points that deviated from the `power-law' scaling (Fig.~4c). All the scattered points were below the fitting (dotted line in Fig.~4c), indicating that, compared with the normal compact excitations (Fig.~4d, e), the morphology of excitations corresponding to the scattered ones were more fractal (Fig.~4f). Moreover, the fractal excitations were much more likely to occur for $\phi$ close to $\phi_\textrm{MCT}$ (Figs.~4c, S8), in accordance with the theoretical prediction of RFOT \cite{2006NP_Stevenson}.

\vspace{12pt}
\textbf{Summary and perspective}

To brief our experiments in one sentence, the dynamical responses of a glassy material was the outcome of the interplay between the external driving and the structure of the material itself (Figs.~3, 4, S9, S10). DH was proved to be strongly encoded with LS, in both linear and non-linear regions (Figs.~1, S5, S6, S7). A non-monotonic DLS (Figs.~3a, S10a) and degree of heterogeneity (Figs.~3b, S10a) was observed in non-linear region. In addition, a universal scaling existed between the $S_\textrm{max}$ and $R_\textrm{g}$ of the excitations (Fig.~4c), while the morphology heterogeneity occurred at middle length scale (inset of Fig.~4c). These results should be general for systems other than colloids since they were controlled by simple factors, i.e., the strength of the driving force and the local structure of the materials. We expect our results to be confirmed in other experimental systems like polymers, metallic glasses, jammed granular as well as in simulations.

It is worthwhile to quantify more on the relaxation of HMR. Both the morphology and the microscopic dynamics of the excitations are probably good distinguisher of linear and non-linear regions. Investigations on how the momentum was transferred after the external perturbation was imposed, especially for the cases where the excitation was fractal, will offer clear experimental observations for the test of theories on the relaxation of glasses \cite{2011RMP_Berthier}. By analyzing the decaying form of the excitations' spatial correlation, it's possible to reveal whether the scale free relaxation \cite{2014PRL_Lemaitre} exists in glassy materials.  

The out-of-phase behavior of the driving force (Figs.~1b, S2a, displacement of particle closest to the laser spot) and the response (Fig.~S2b, time evolution of $S_\textrm{max}$ as a joint consequence of both inner and outer particles) raised the interesting problem of loss during the excitation (See discussion in SI); although the $S_\textrm{max}$ we used in this manuscript was frequency independent \cite{2012RPP_Hunter}. For example, we observed severer time lag between $S$ (the response) (Fig.~S2b) and the displacement of nearby particles to the laser spot (the driving force) (Fig.~S2a), implying higher loss in non-linear region. This observation was in accordance with the non-affine or plastic feature in non-linear region ($\phi=0.65$ curve in Fig.~S2a). Coining the strategy in the shear experiment \cite{2009N_Mattsson}, it could be interesting to test both the storage and loss curves of $S$ upon a series of perturbation with time separation much shorter than the relaxation time. Such experiments will offer insight into the hysteresis and memory properties of glassy materials.

The experimental strategy in this manuscript paves the way of investigating a broad range of non-equilibrium phenomena in material science, geology and biophysics. By actively inducing local excitations, we are able to directly reveal the causal relation between perturbation the response on the microscopic level. This strategy is therefore particularly suitable for the study of processes like avalanche, earthquake and friction \cite{2012RMP_Kawamura}, in which topological excitation exists. By controlling the motion of a few critical leaders, one can also investigate how individual behaviors influence the group decisions of animal flocks \cite{2017NP_Reichhardt}, and thus offer better understanding to the general rules that controls complex \cite{2016RMP_Liu} and living \cite{2018RMP_Munoz} systems.

\vspace{12pt}
\textbf{Methods}

In this experiment, we mixed $\sigma_\textrm{s}=(2.08 \pm 0.05)~\mu m$ and $\sigma_\textrm{l}=(2.91 \pm 0.05)~\mu m$ PMMA colloidal particles (Microparticle GmbH) homogeneously with the number ratio of $0.55:0.45$, according to the Kob-Anderson model \cite{1995PRE_Kob}. This binary spherical system was able to avoid crystallization effectively and hence served as good glass formers \cite{2018PRL_Pedersen}. A small amount ($<$0.1$\%$ in number) of 3.37$~\mu m$ gold particles (Microparticle GmbH) was added to the suspension. They bore dual functionality of both spacers that avoided stuck of 2.91$~\mu m$ PMMA colloidal particles and heat absorber in the laser excitation experiments. To maximize the xy-component of the scattering force, we adopted a quasi$\textrm{-}$two$\textrm{-}$dimensional sample configuration, in which colloidal particles were dispersed in water between two parallel horizontal glass walls, the separation between which was less than 1.2~$\sigma$, screening off the majority of the hydrodynamic force between particles \cite{2013PRL_Ma}. For colloids in 2D, the area fraction $\phi=\frac{1}{4} \pi (\sigma_\textrm{s}^{2} \rho_\textrm{s}+\sigma_\textrm{l}^{2} \rho_\textrm{l})$ played the same role as the inverse temperature $1/T$ plays in molecular systems, with $\rho_\textrm{s}$ and $\rho_\textrm{l}$ the number density of small and large particles respectively. Totally eight area fractions within $0.45\leq\phi\leq 0.80$ were recorded through video microscopy. During the 2- to 6-$hours$ measurements at each $\phi$, no drift, flow, or density change were observed. The center-of-mass positions of the colloidal particles were tracked using an image-processing algorithm \cite{1996ARPC_Murray}.

Two laser excitation methods were adopted in our experiments. In the `gold-particle' method, a single laser beam was shined to the center of a doped gold particle \cite{2017PNAS_Buttinoni}. The gold particle was able to absorb heat efficiently (Fig.~S1b) and thus enhanced the mobility of its neighboring particles. In the `four-point' method, we used a spatial light modulator to split one beam into four and simultaneously shined these sub-beams to different positions (without gold particles) of the sample (Fig.~S1a). In this case, the scattering force (Fig.~S1c) was the dominant effect that the laser beam exerted to colloids. When a beam of laser was shined into the sample, it will exert a force to the particles near the edge of the laser waist, pushing them away from the beam center. When the colloidal particles were strictly confined between two glass walls, the pushing force in z-direction will be able to generate xy-component due to the collisions between the colloidal particles and the glass wall. Moreover, the scattering force was proved to be localized and the magnitude of which was proportional to $A$ \cite{2015OE_Villadsen, 2016APhot_Zensen}. The spatial separation between light spots was more than 20~$\sigma_\textrm{s}$, much larger than the DLS in the deepest non-linear region. This arrangement avoided the interference between excitations and thus ensured the statistical independence of each event. The time separation between two excitations was much larger than the relaxation time of the glass sample at corresponding $\phi$.

The excitation size, $S_\textrm{max}$, was quantified by the following steps. For each frame, we selected the particles whose displacement (within $\Delta t=5~s$) exceeded a threshold value as the excited particles, and used the total area of these particles as $S$ at this time. The threshold was chosen as $0.3~\sigma_\textrm{s}$ basing on the Lindemann's melting criteria \cite{1910PZ_Lindemann}. After the laser was off at $t=0.5~s$, the $S$ kept increasing with $t$ and reached a maximum at about $t=1.5~s$. We used this maximum value of $S$ as the excitation size under that ($\phi,~A$). The trend of $S_\textrm{max}$ (Fig.~S4a) and $\sigma(S_\textrm{max})$ (Fig.~S4b) as a function of $\phi$ remained the same by replacing the global $\phi$ by local $\phi$, excluding the possibly influence by density fluctuation. For each ($\phi,~A$), more than 20 individual excitation events were evaluated for reliable statistics.

\vspace{12pt}
\textbf{Supplementary Information} is at the end of this document.

\vspace{12pt}
\textbf{Acknowledgements} This study was supported by Korean Institute for Basic Science (project code IBS-R020-D1).

\newpage
\section{Supplementary Information for ``Probing the Local Response of Glass-forming Liquids by Laser Excitations''}

\vspace{12pt}
\textbf{The experimental system}

In our experiments, one or four local laser beams of 1-2~$\sigma_\textrm{s}$ in width and 0.5~$s$ in time (Fig.~S1a) was imposed to the sample, enhancing the mobility of nearby colloidal particles. We named this process in the main text as an excitation event. We found there were two major physical mechanisms contributing to the excitation event: the heating of local environment (Fig.~S1b) and the scattering forces (Fig.~S1c). Both the two effects were able to be established within 0.5~$s$ after the laser was on and disappear immediately when the laser was off; verifying them as exclusive mechanisms that caused the HMR. Other effects like the convection flow \cite{2018NCB_Mittasch} and thermophoresis of colloidal particles \cite{2009PRL_Jiang} took place at much longer time scales than our pulse duration; excluding the possibility that they played a role in the excitation. Moreover, both the heating and scattering force effects were local in space, roughly the same size of the laser beam ($< 2~\sigma$).

Heating was the major effect in the `gold-particle' method. As described in \textbf{Methods} section of the main text, since metal was able to absorb heat much more efficiently than water or PMMA. To quantify the heating effect, we added a thermal sensitive fluorescein (BCECF, Thermo Fisher Scientific) into DI water and plotted the brightness profile across the light spot 0.5~$s$ after the laser was on. It was reported in ref.~\cite{2002PRL_Braun} that the brightness of this fluorescein was enhanced by three times when the temperature was 6~$^{\circ}$C higher than the ambient temperature. Since the emitted luminescence of the fluorescein linearly increased with $T$, we then obtained the local temperature ($T$) at the beam center by measuring the peak values of the brightness profile. As expected, $T$ increased linearly with $A$ (Fig.~S1b (inset)). More importantly, the broadness of the brightness profile did not significantly increase with $A$, indicating the region influenced by the heating effect was roughly the size of the laser beam, i.e., less than 2~$\sigma$ (Fig.~S1a (inset)). The reliability of this method within colloidal system has been verified by previous melting \cite{2012S_Wang} and thermophoresis \cite{2009PRL_Jiang} experiments.

In the `four-point' method, the scattering force was the dominant factor. As a type of optical forces, the scattering force represents the momentum transfer from the external radiation field to the colloidal particle by scattering and absorption. When a short pulse was imposed to colloidal particles, it tends to `kick' them away from the center of the beam \cite{2015OE_Villadsen, 2016APhot_Zensen}. Previous studies showed the magnitude of the scattering force also linearly increased with $A$ (Fig.~S1b) \cite{2016APhot_Zensen}; and its influencing range was roughly the size of the laser beam \cite{2015OE_Villadsen}. Although the scattering force was in z-direction, in our case of 2D confined system, the particle collided with the glass substrate frequently and thus generated significant xy-component of the force. 

The out-of-phase behaviors of particle displacements (Fig.~S2a) and excitation size (Fig.~S2b) as a function of time enabled our experiments revealing similar information to that from shear experiments, where frequency dependent storage and loss were measured. In our experiment, the sudden movements of the particles closest to the laser spot initialized the collision with outer layered particles and thus caused the relaxation of HMR (Fig.~S2c-e). And since the displacements of such particles reached the maximum within an extremely short period ($0.3~s$) after the laser was on, this sudden position changes at $t=0~s$ (Figs.~1a, b, S1a) could be regarded as a $\delta-$function-like driving force. The size of excitation as a function of time was then the response (Fig.~S1b). In the experiment described in the main text, we used the peak value, $S_\textrm{max}$, to describe the excitation size under given ($\phi,~A$). This quantity was in principle frequency independent and was always reached after the laser was off at $t=0.5~s$. Note that the maximum value of the particle displacement in Figs.~1a, S2a was much smaller than the dimension of the excitation events, proving that the perturbation in our experiment was truly local.

More quantification of dynamics and structure demonstrated the system was well in equilibrium before each excitation. The $\Delta t=5~s$ we adopted for the calculation of $S_\textrm{max}$ and $\sigma(S_\textrm{max})$ was much smaller than the relaxation time measured from the intermediate scattering function (Fig.~S3c), which was further much smaller than the time separation between two excitation events. These choices ensured no crossing time scales that can cause artificial non-monotonic behaviors similar to that in Fig.~3. The similar shape of radial distribution function ($g(r)$) for all samples with $\phi$ ranging from $0.451$ to $0.826$ (Fig~S3d) indicated that no structural instability within our experiments. 

The effect of density fluctuation in space was evaluated in Fig.~S4. By replacing the global $\phi$ with local $\phi$ (area fraction within 2 $\sigma_\textrm{l}$ distance to the laser spot), we found the non-monotonic dynamical length scale did not change (Fig.~S4), although the variance of $S_\textrm{max}$ was significantly depressed. Since in colloidal system, $\phi$ played the same role as thermodynamic temperature; and since $S_\textrm{max}$ was determined by both the LS near the laser spot and those on the relaxation passage of HMR, the global $\phi$ adopted in the main text seemed to be a more reasonable thermodynamic variable in our experiment.

\vspace{12pt}
\textbf{More evidences to the connection between dynamics and structure}

As a complement to the `gold particle' method in the main text (Fig.~2), we conducted experiments using the `four-point' method in the linear region. When the laser beam was weak, no exchange of positions among particles would occur. After a slight enhancement of rattling inside the cage, the fast moving particles soon relaxed back to their equilibrium positions when the excitation was over. Therefore, the LS at a fixed shining spot was exactly the same before each excitation event. Again, excitation patterns for the same spot but at different trials were very similar to each other for both low laser power cases (Figs.~S5, S6).

An advantage of `four-point' method is that we were able to compare the excitation patterns at positions with distinct LS simultaneously. As clearly shown in Figs.~S5, S6, the morphology of the patterns varied from spot to spot. It therefore offered a strong collateral evidence to the conclusion we obtained in the main text: DH was deeply encoded with LS.

The link between dynamics and structures held in both linear and non-linear regions. In Fig.~S7a, we observed a positive correlation between $S_\textrm{max}$ and local elastic modulus ($\sigma_\textrm{xy}$). Here, $\sigma_\textrm{xy}$ was measured by `stress assessment from local structural anisotropy (SALSA)' method \cite{2016NM_Lin}, which was proved to be highly accurate for quantifying the elasticity of liquids \cite{2017PRL_Lin} and crystals \cite{2016NM_Lin} composed of hard-sphere-like colloidal particles. Regardless of $\phi$ and $A$, the $S_\textrm{max}$ was solely determined by the local elasticity of the position at which the external perturbation was imposed. We know that $S_\textrm{max}$ reflected the extent to which the relaxation of a HMR can reach and thus was an indicator of DH; while according to SALSA method, $\sigma_\textrm{xy}$ was eventually determined by the LS. Therefore, the positive correlation between $S_\textrm{max}$ and $\sigma_\textrm{xy}$ was a direct link between dynamics and structure. Such positive correlation could be weakly but directly verified by the spatial mapping between particles' mean square displacements (Fig.~S7b) and $\sigma_\textrm{xy}$ (Fig.~S7c). In addition to the size of excitation, the morphology of the excitation pattern was able to be spatially mapped with the local elasticity field; implying the relaxation path of the HMR was influenced by the elasticity field around the shined spot. In another word, the momentum of high mobile particles tended to be transferred following the `weak' passage, along which the local elasticity was small. The fact that the data points in Fig.~S7a included excitation events in both linear and non-linear regions generalized the conclusion that DH was encoded with LS to the whole $\phi$~-~$A$ phase diagram of colloidal glass. 

\vspace{12pt}
\textbf{From linear to non-linear region}

When $A$ was gradually increased, the system was driven from linear to non-linear region. The threshold between linear and non-linear regions was determined by the fluctuation rather than thermodynamic average. In Fig.~4a, b, we plotted $S$ and $\sigma(S_\textrm{max})$ as a function of $A$ under different fixed $\phi$ values. As expected, both the DLS and the DH were enhanced under increasing driving force (Fig.~S8). However, it was $\sigma(S_\textrm{max})$ (Fig.~S8b) instead of $S_\textrm{max}$ (Fig.~S8a) that exhibited a clear deviation from linear form when $A$ became large. The absence of non-linear increasing of DLS in the whole $A$ range (Fig.~S8a) implied that the average deformation of a glassy material is primarily determined by the driving force. In this sense, the non-linearity was a fluctuation determined quantity (Fig.~8b), which reflected the heterogeneity of the glassy materials and was irrelevant to the thermodynamic averaging. Fig.~S8 showed the excitation patterns under increasing $A$. It could be directly observed that both the size and morphology of the excitation became progressively diverse when the system went deeper and deeper into non-linear region (Figs.~4d-i, S8c-e); corresponding to the non-linear increasing of $\sigma(S_\textrm{max})$ in Fig.~S8b. However, although the $\sigma(S_\textrm{max})$ deviated from linear increasing form at large $A$, $\sigma(S_\textrm{max})(A)$ exhibited no sudden jump, indicating a rather broad and continuous transition zone between linear and non-linear regions.

A natural question extended from the $\sigma(S_\textrm{max})(A)$ plot (Fig.~S8b) was how to more clearly distinguish the non-linear responses from linear ones. Here we offer two clues for follow-up studies. One thing is to take a closer look at the relaxation of the HMR. In Fig.~S2e, we marked (by the red crosses) all particles that underwent irreversible rearrangements for the left-down corner excitation event. In contrast to the slight rattling within cages in linear events, particle exchanged positions much more frequently in non-linear excitations. Could the rate of irreversible rearrangement be a clearer distinguisher that defined the boundary between linear and non-linear regions deserves more exploration. The second feature of non-linear region that worth mentioning was the altering in the form of probability distribution function ($P(S_\textrm{max})$) of the excitation size (Fig.~S9a). At a fixed high laser power, the $P(S_\textrm{max})$ in non-linear region ($\phi=0.60$) qualitatively differed from that in linear region ($\phi=0.51$). More excitations with large $S_\textrm{max}$ occurred when the system was driven into non-linear region, namely, the peak in the blue curve of Fig.~3a in this example. It's also worth noting that the non-monotonic behaviors disappeared (Figs.~3, S9b) when the $A$ was too small to drive the system into non-linear region.

In addition, preliminary data in Fig.~S2a, b suggested that the time lag between the response and the driving force was most prominent in middle $\phi$ region. At the same $\phi=0.65$, the particle displacement exhibited clear non-affine or plastic feature ($\phi=0.65$ curve in Fig.~S2a), which was caused by vast particle rearrangement (Movie 2) in the non-linear region. For fixed very high laser power, the non-monotonic time lag as a function of $\phi$ seemed to be in accordance with the non-monotonic dynamical length scale reported in the main text (Fig.~3a, b). 

\vspace{12pt}
\textbf{The $\phi-A$ `linearity' phase diagram}

The non-monotonic $S_\textrm{max}$ and $\sigma (S_\textrm{max})$ reflected a non-trivial competition between collision frequency and rigidity during the process in which the elasticity was established. When $\phi$ increased under a fixed $A$ value, two competing effects existed within the system. On one hand, the average distance among particles decreased with $\phi$. The smaller distance of a particle to its neighbors made the collisions easier to occur \cite{1996N_Dzugutov}. It was shown by liquid theory that the peak value of $g(r)$ is proportional to the collision rate \cite{1996N_Dzugutov}. The fact that the value of $g(r=r_\textrm{first peak})$ almost doubled (Fig.~S3d inset) when $\phi$ increased from 0.45 to 0.83 indicated significant enhance of collision rate in high $\phi$ region. The higher collision rate then enhanced the cooperative motion that eventually facilitated the relaxation (Movie 2). In this sense, the driving force had an increasingly stronger impact on the dynamics. On the other hand, the rigidity was gradually established when the system went from liquid to glass region. As the rigidity of a material substantially increasing near the glass transition point ($\phi_\textrm{g}$) \cite{2015PRX_Klix, 2016PRL_Saw}, particles were so strongly caged by their neighbors that they could hardly move under the same driving force, i.e. fixed $A$ value (Movie 3). In another world, the higher rigidity weakened the cooperative motion and thus depressed the relaxation. Therefore, these two competing effects together caused the non-monotonic $S_\textrm{max}$ as a function of $\phi$. Since cooperativity also strongly influences the frequency of irreversible rearrangement, which determines DH of a material, this scenario also explains the non-monotonic $\phi$-dependence of $\sigma (S_\textrm{max})$.

Integrating all $S(\phi, A)$ data points into a 3D plot (Fig.~S10a), we were able to construct a schematic of the $\phi-A$ `linearity' phase diagram (Fig.~S11a). The winding phase boundary between linear (blue `L' in Fig.~S10b) and non-linear (red `N' in Fig.~S10b) regions came from the intriguing interaction between the external perturbation and the glassy material itself; and it offered a consistent explanation on both $A$-dependent $\phi_\textrm{max}$ and $\phi$-dependent $A_\textrm{TH}$. In non-linear region, the high chance of irreversible rearrangements will naturally lead to larger DLS as well as the heterogeneity of it. When $A$ was small, the driving force was too weak to cause particles colliding with their neighbors, not to mention exchanging of positions (Movie 1). The excitation just involved some rattling of particles near the laser spot that were independent of $\phi$. Therefore the system was in linear region for all $\phi$, i.e., no peak in $S_\textrm{max}(\phi)$ plot (low laser power curve in Figs.~3a). However, when we fixed $A$ to a fairly larger value, in certain $\phi$ region, the competition between the higher collision frequency (caused by decreasing of inter-particle distance) and stronger caging effect (due to enhancing rigidity) reached a favorable balance, defining the non-linear region with both the largest DLS and DH.

In brief, one general and applicable guidance that our experiments can provide is that a glassy material is most sensitive to external perturbations in middle temperature region, where the elasticity is initially established. In this region, the collective motion among motifs reaches a maximum because the vibration of atoms or molecules is not yet undermined by the caging effect. Due to the simplicity of the controlling factors in our experiments, we expect this non-monotonic DLS to be observed in other experimental systems as well as simulations.

\renewcommand{\thefigure}{S1}
\begin{figure}[!h]
	\centering
	\includegraphics[width=1.0\columnwidth]{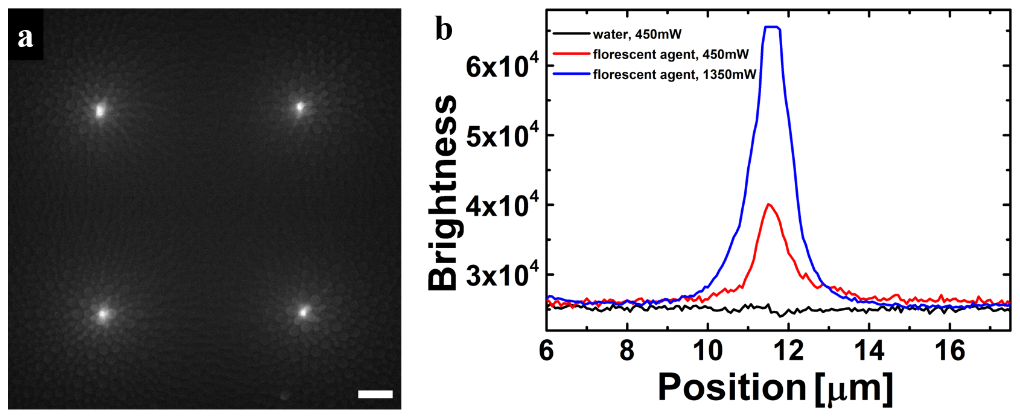}
	\caption{\textbf{a}, The bright field image of a sample under four-point laser excitation. \textbf{b}, The brightness profile across a laser spot in a fluorescein sample. \textbf{c}, The scattering force as a function of $A$.}
	\label{fig:S1}
\end{figure}

\renewcommand{\thefigure}{S2}
\begin{figure}[!h]
	\centering
	\includegraphics[width=1.0\columnwidth]{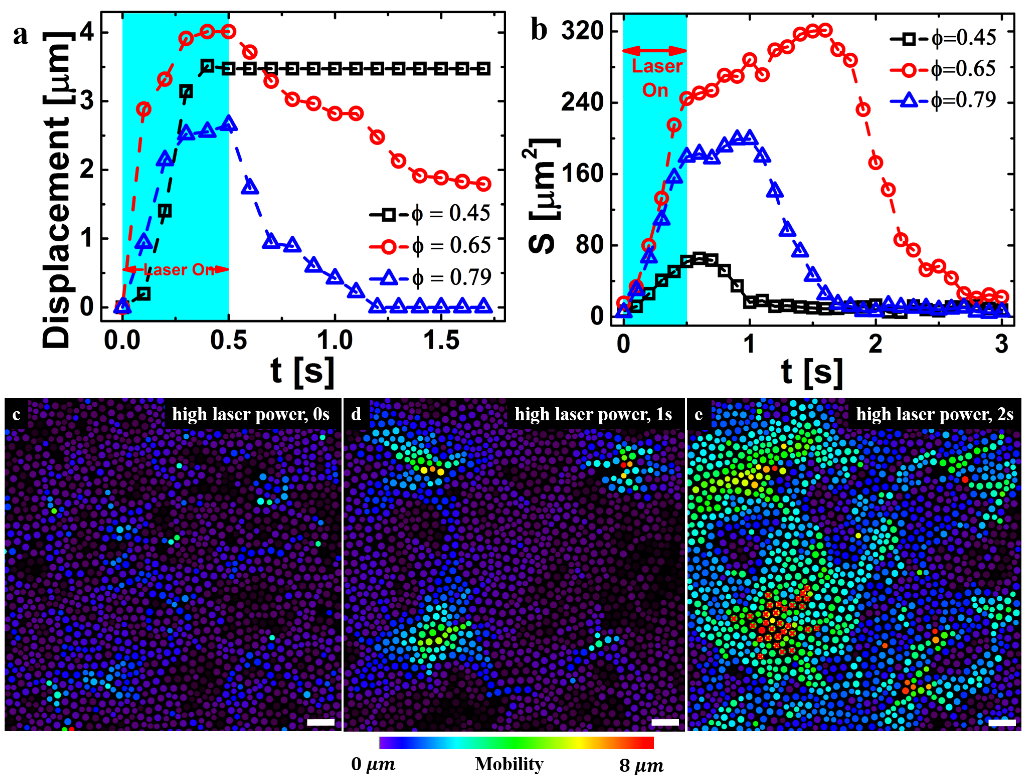}
	\caption{\textbf{a}, Displacements of particles closest to the laser spot of high laser power as a function of time. \textbf{b}, Size of an excitation as a function of time under high power laser beam. The duration of the laser was 0.5~$s$. \textbf{c-e}, Excitation patterns under high laser power four-point excitation in a $\phi=0.70$ sample after \textbf{c}, 0~$s$, \textbf{d}, 1~$s$, \textbf{e}, 2~$s$ the laser was off. The color represents the particles' displacements within 5~$s$. The duration of the pulse was 0.5~$s$. The scale bar is 10~$\mu m$.}
	\label{fig:S2}
\end{figure}

\renewcommand{\thefigure}{S3}
\begin{figure}[!h]
	\centering
	\includegraphics[width=1.0\columnwidth]{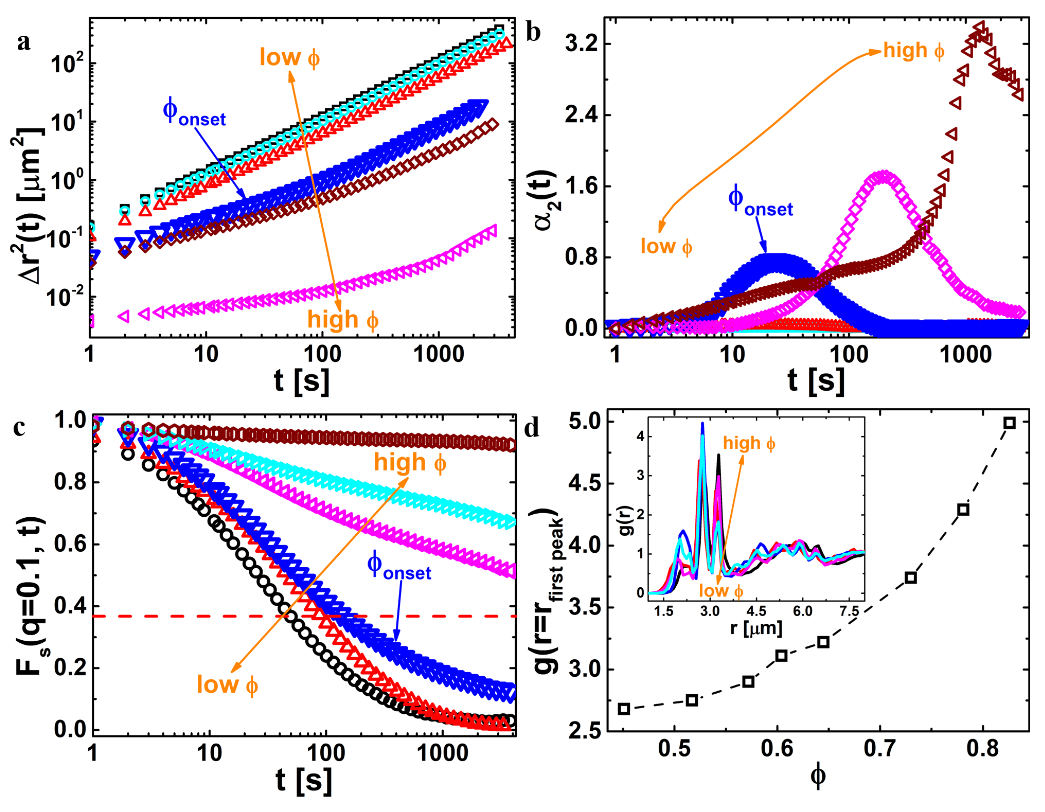}
	\caption{\textbf{a}, The mean square displacement as a function of time for samples with different $\phi$. \textbf{b}, The non-Gaussian parameter as a function of time for samples with different $\phi$. \textbf{c}, Intermediate scattering function for samples with different $\phi$. \textbf{d}, The radial distribution function ($g(r)$) for samples with different $\phi$. \textbf{Inset,} The first peak value of $g(r)$ as a function of $\phi$.}
	\label{fig:S3}
\end{figure}

\renewcommand{\thefigure}{S4}
\begin{figure}[!h]
	\centering
	\includegraphics[width=1.0\columnwidth]{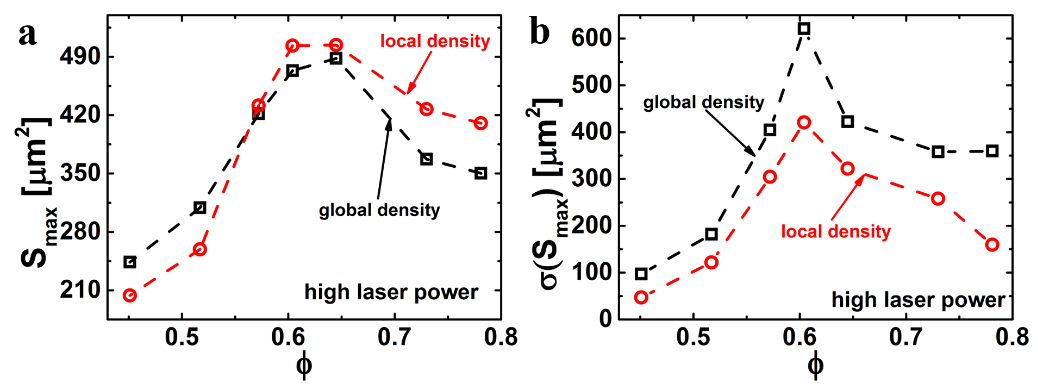}
	\caption{\textbf{a}, Excitation size as a function of $\phi$ and local $\phi$ under a fixed high laser power. \textbf{b}, Variance of excitation size as a function of $\phi$ and local $\phi$ under a fixed high laser power.}
	\label{fig:S4}
\end{figure}

\renewcommand{\thefigure}{S5}
\begin{figure}[!h]
	\centering
	\includegraphics[width=1.0\columnwidth]{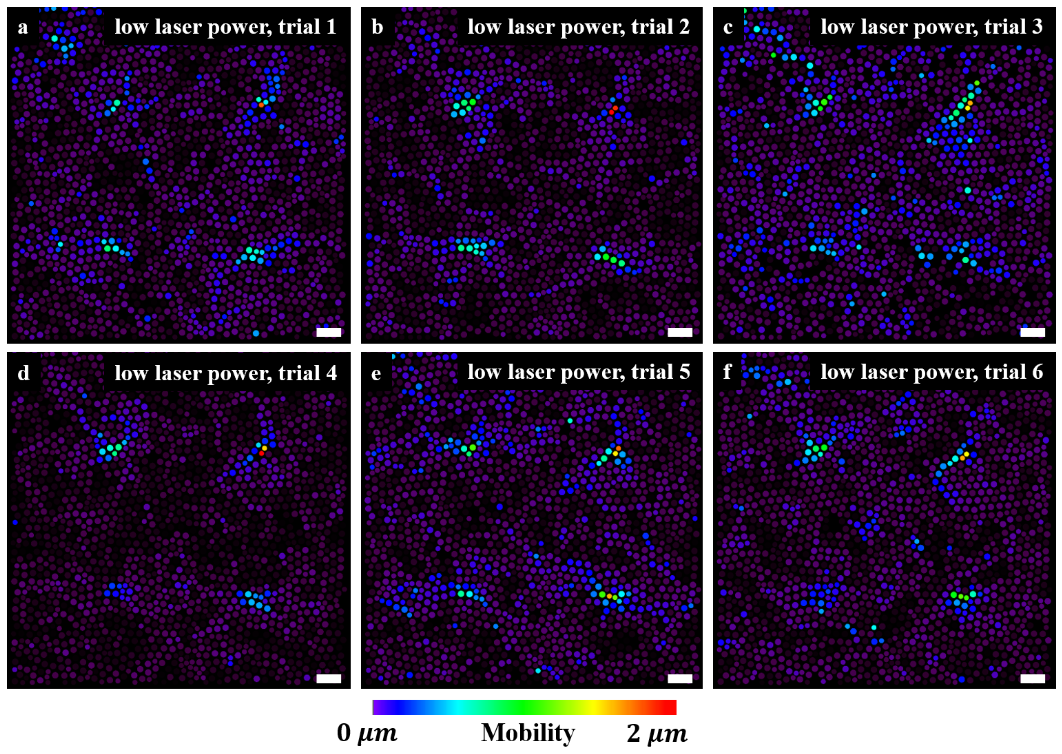}
	\caption{Excitation patterns under low laser power four-point excitation in a $\phi=0.58$ sample. The color represents the particles' displacements within 5~$s$ at $t=0.9~s$. The duration of the pulse was 0.5~$s$. The separation between two excitation events was 20~$min$. The scale bar is 10~$\mu m$.}
	\label{fig:S5}
\end{figure}

\renewcommand{\thefigure}{S6}
\begin{figure}[!h]
	\centering
	\includegraphics[width=1.0\columnwidth]{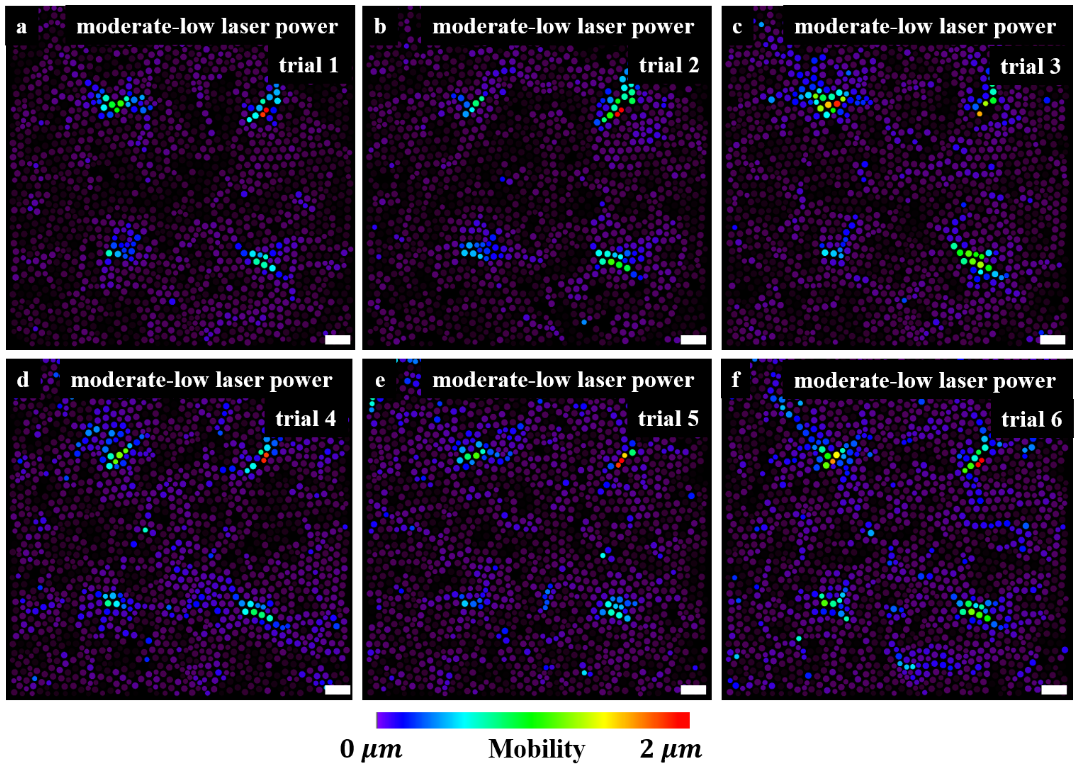}
	\caption{Excitation patterns under moderately low laser power four-point excitation in a $\phi=0.58$ sample. The color represents the particles' displacements within 5~$s$ at $t=1.0~s$. The duration of the laser pulse was 0.5~$s$. The separation between two excitation events was 20~$min$. The scale bar is 10~$\mu m$.}	
	\label{fig:S6}
\end{figure}

\renewcommand{\thefigure}{S7}
\begin{figure}[!h]
	\centering
	\includegraphics[width=1.0\columnwidth]{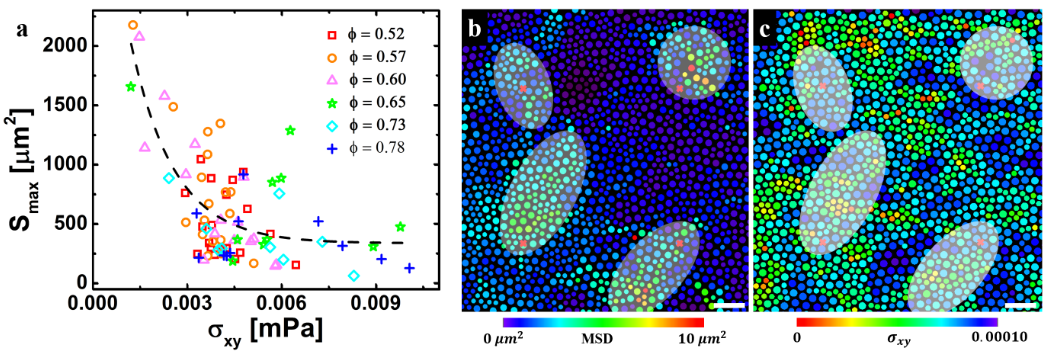}
	\caption{\textbf{a}, Excitation size as a function of local elastic modulus ($\sigma_\textrm{xy}$). The symbol of the data points represents the area fraction, $\phi$. The laser is fixed to a high power for all the data points. \textbf{b, c,} The spatial mapping of dynamical relaxation and local elasticity in a $\phi=0.65$ sample. The color in \textbf{a} represents particles' mean square displacement within 20~$s$ after the laser was off. The color in \textbf{b} represents particles' local elastic modulus, $\sigma_\textrm{xy}$, calculated based on structures. The duration of the laser was 0.5~$s$. The scale bar is 10~$\mu m$.}
	\label{fig:S7}
\end{figure} 

\renewcommand{\thefigure}{S8}
\begin{figure}[!h]
	\centering
	\includegraphics[width=1.0\columnwidth]{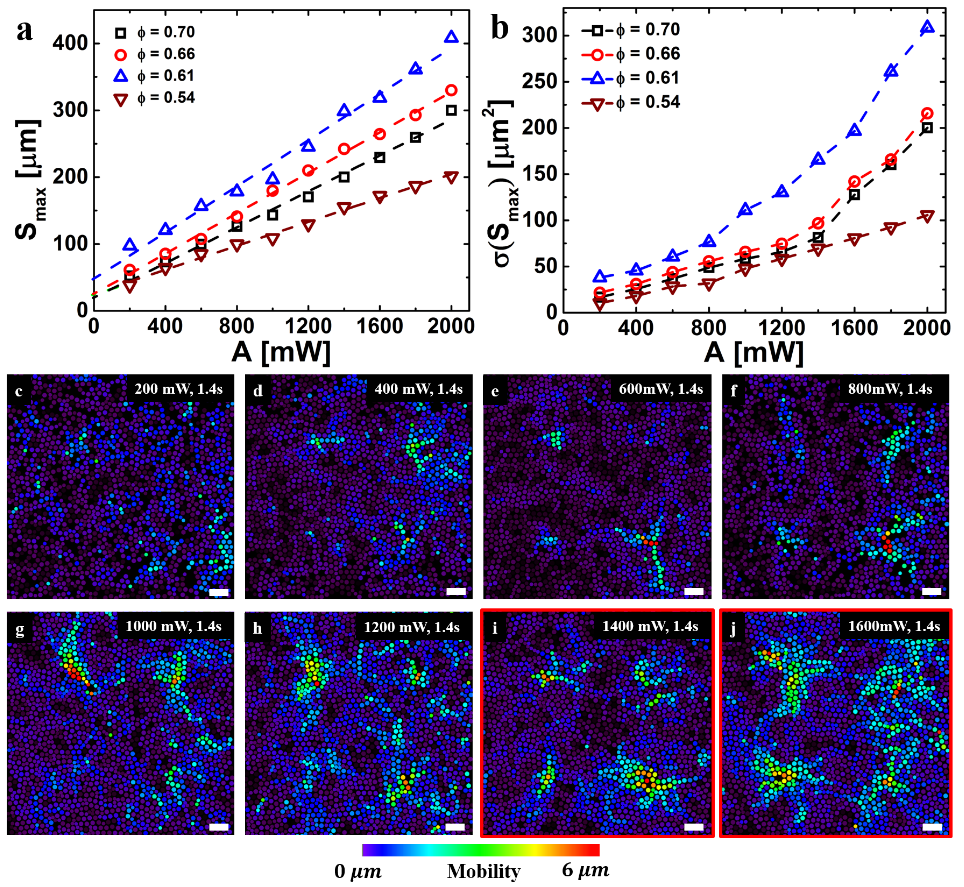}
	\caption{\textbf{a}, The averaged excitation size as a function of $A$. \textbf{b}, The averaged variance of excitation size as a function of $A$. \textbf{c-j}, Excitation patterns from linear to nonlinear region. Four same laser pulses with increasing $A$ were simultaneously shined to a $\phi=0.70$ sample. The color represents the particles' displacements within 5~$s$ at $t=1.2~s$. The duration of the pulse was 0.5~$s$. The separation between two excitation events was 20~$min$. The scale bar is 10~$\mu m$.}
	\label{fig:S8}
\end{figure}

\renewcommand{\thefigure}{S9}
\begin{figure}[!h]
	\centering
	\includegraphics[width=1.0\columnwidth]{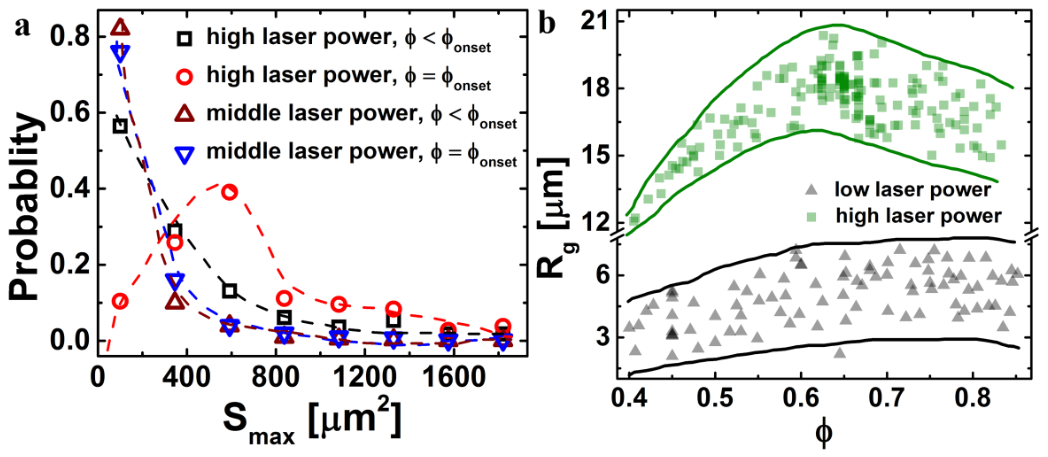}
	\caption{\textbf{a}, The probability distribution function of $S$ in linear ($\phi=0.51$) and non-linear regions ($\phi=0.60$). Both curves were obtained on the bases of 40 excitation events. \textbf{b}, The radius of gyration ($R_\textrm{g}$) as a function of $\phi$ for a fixed lower laser power and a fixed high laser power.}
	\label{fig:S9}
\end{figure}

\renewcommand{\thefigure}{S10}
\begin{figure}[!h]
	\centering
	\includegraphics[width=1.0\columnwidth]{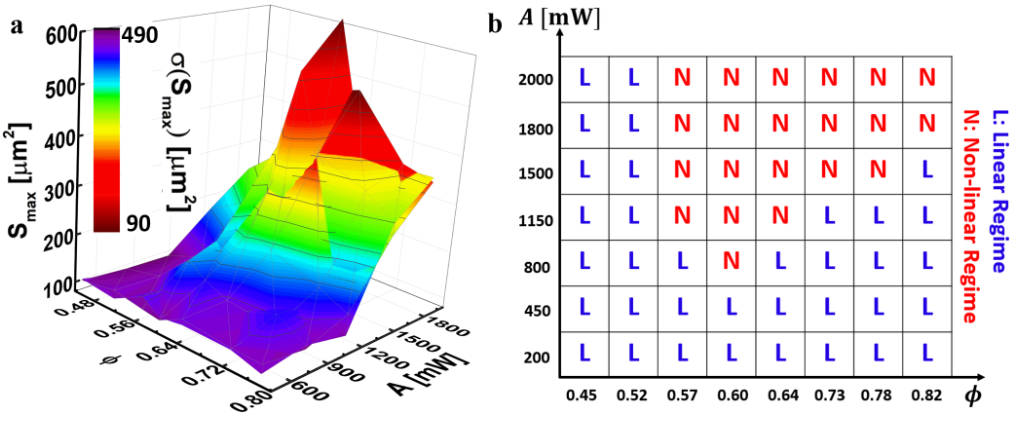}
	\caption{\textbf{a}, The 3D-plot of excitation size as a function of both $\phi$ and $A$. The color represents the variance of the excitation size. \textbf{b}, A schematic of the $\phi-A$ linearity `phase diagram' with separated linear and non-linear region.}
	\label{fig:S10}
\end{figure}

\end{document}